\documentclass[12pt,preprint]{aastex}

\newcommand{\be}{\begin{equation}}
\newcommand{\ee}{\end{equation}}
\newcommand{\mdot}{{\stackrel{\cdot}{M}}}

\begin{document}

\title{GRAVITATIONAL COLLAPSE OF MAGNETIZED CLOUDS
\\ I. IDEAL MHD ACCRETION FLOW}

\author{Daniele Galli}
\affil{INAF-Osservatorio Astrofisico di Arcetri \\
              Largo Enrico Fermi 5, I-50125 Firenze, Italy \\
              galli@arcetri.astro.it}

\author{Susana Lizano}
\affil{Centro de Radioastronom\'{\i}a y Astrof\'{\i}sica, UNAM \\
              Apdo. Postal 3-72, Morelia, Michoac\'an 58089, Mexico\\
              s.lizano@astrosmo.unam.mx}

\author{Frank H. Shu}
\affil{Physics Department, National Tsing Hua University \\
              Hsinchu 30013, Taiwan, Republic of China \\
              shu@mx.nthu.edu.tw}

\author{Anthony Allen}
\affil{Institute of Astronomy and Astrophysics \\
              Academia Sinica, Taipei 106, Republic of China \\
              tony@asiaa.sinica.edu.tw}

\begin{abstract}
We study the self-similar collapse of an isothermal magnetized rotating
cloud in the ideal magnetohydrodynamic (MHD) regime. In the limit of
small {distance from the accreting protostar} we find an analytic
solution that corresponds to free-fall onto a central mass point. The
density distribution is not spherically symmetric but depends on the
mass loading of {magnetic} field lines, which can be obtained by
matching our inner solution to an outer collapse solution previously
computed by Allen, Shu \& Li. {The concentration of magnetic field
trapped by the central mass point under field-freezing, independent on
the details of the starting state, creates a split monopole
configuration where the magnetic field strength increases as the
inverse square of the distance from the center. Under such conditions,
the inflow eventually becomes subalfv\`enic and the outward transfer of
angular momentum by magnetic braking very efficient, thus preventing
the formation of a centrifugally supported disk.} Instead, the
azimuthal velocity of the infalling gas decreases to zero at the 
{center}, and the gas spirals into the star.  Therefore, the dissipation
of dynamically important levels of magnetic field is a fundamental
requisite for the formation of protoplanetary disks around young
stars. 
\end{abstract}


\section{Introduction}

Since the realization that magnetic fields exist in interstellar
clouds, there has been the potential for a severe magnetic-flux problem
in star formation.  If field freezing were to apply (the ideal MHD 
{conditions}), newly formed solar-type stars should have surface fields
of $\sim 10^7$~G (e.g., Mestel \& Spitzer 1956), which is several
thousand times larger than given by the observations of T Tauri stars
(e.g., Johns-Krull, Valenti, \& Saar 2004).  Magnetic-field dissipation
is crucial to the resolution of this contradiction. In this paper, we
demonstrate that it is also needed to avoid an extremely efficient
magnetic braking that would prevent the formation of the observed
protoplanetary disks around young stars.

Allen, Shu \& Li (2003; hereafter A03a) performed informative numerical
simulations of the collapse of isothermal magnetized toroids, {which are
axially symmetric equilibrium configurations with spatially uniform
mass-to-flux ratio (such configuration is called {\it isopedic}).  These are
pivotal states at time $t=0$, in the sense that they represent the
idealized state of a molecular cloud core at the instant of formation
of a central density cusp, and separate the pre-protostellar phase of
evolution from the protostellar accretion phase (Li \& Shu~1996). The
structure of these toroids is completely determined by the single
parameter $H_0$, that is the fraction of support provided against
self-gravity by poloidal magnetic fields relative to that of gas
pressure (Li \& Shu 1996).  The collapse of singular isothermal toroids
occurs {inside-out} in a self-similar manner (A03a), in which
matter collapses toward the central point mass dragging magnetic field
lines along with it}.

Later, Allen, Li \& Shu (2003; hereafter A03b) included angular
momentum characterized by a flat rotation curve in these models.  The
formation of a centrifugally supported disk around the central star did
not occur for any reasonable nonzero value of the magnetic support
parameter $H_0$. They concluded that, in ideal MHD conditions (i.e., no
magnetic field dissipation), the efficiency of magnetic braking was so
high as to prevent disk formation.  Simulations with limited numerical
resolution can be suggestive, but proof requires an analytic or
semi-analytic demonstration.  Analytic or semi-analytic results have
the added advantage that they can be the starting points for additional
development, such as providing a background equatorial collapse against
which a bipolar outflow might simultaneously be taking place along the
poles (Lizano et al.~2006, in preparation), or for an extension into the
non-ideal regime (Shu et al.~2006).

Motivated by these considerations, in this paper we {determine
analytically} the asymptotic behavior of the density, velocity and
magnetic field in the self-similar collapse of an isothermal magnetized
rotating cloud.  We find that {at small scales the density and
poloidal velocity field in the collapse solution approaches free-fall
on a split magnetic monopole configuration}, with the azimuthal
velocity decreasing to zero at the origin.  This result proves that,
without magnetic field dissipation, a centrifugally supported disk {
around the accreting central star} cannot form.

The organization of the paper is as follows.  In \S 2 we discuss the
relevant {ideal} MHD equations.  In \S 3 we {put these
equations in non-dimensional self-similar form. In \S 4 we  write the
self-similar equations in a streamline formulation, in terms of the
Grad-Shafranov and Bernoulli's equation, and take the limit for small
values of the self-similar variable $x = r/at$, where $r$ is the spherical
radius, $a$ is the sound speed of the cloud and $t$ is the time
variable. In \S 5 we show some numerical examples of our inner solution
for the case of no rotation, using the numerical simulations of A03a to
obtain an approximation for the mass loading of field lines.  In \S 6
we consider the effects of rotation and show that the azimuthal
velocity vanishes at the origin, i.e., the efficient magnetic braking
prevents the formation of a centrifugally disk around the star.
Finally, in \S 8, we discuss the implications of our work and summarize
our conclusions.}

\section{Ideal MHD equations}

The equations governing the collapse of an isothermal magnetized cloud
are the equation of continuity
\be
{\partial\rho\over \partial t}+\nabla\cdot(\rho{\bf u})=0,
\label{cont}
\ee
where $\rho$ is the density and ${\bf u}$ is the velocity
of the gas; the equation of motion
\be
{\partial {\bf u}\over \partial t}+
(\nabla\times {\bf u})\times {\bf u} + {1\over 2}\nabla |{\bf u}|^2
= -{a^2\over\rho}\nabla\rho-\nabla V+{1\over 4\pi\rho}
(\nabla\times {\bf B})\times {\bf B},
\label{mot}
\ee
where $V$ is the gravitational potential and $ {\bf B}$ is the magnetic
field, the induction equation in the ideal MHD limit
\be
{\partial {\bf B}\over \partial t}  =  \nabla\times({\bf u}\times {\bf B}),
\label{ind}
\ee
the condition of no monopoles,
\be
\nabla\cdot{\bf B}=0,
\label{divB}
\ee
and Poisson's equation
\be
\nabla^2 V=4\pi G\rho,
\ee
where $G$ is the gravitational constant.

\begin{figure}
\plotone{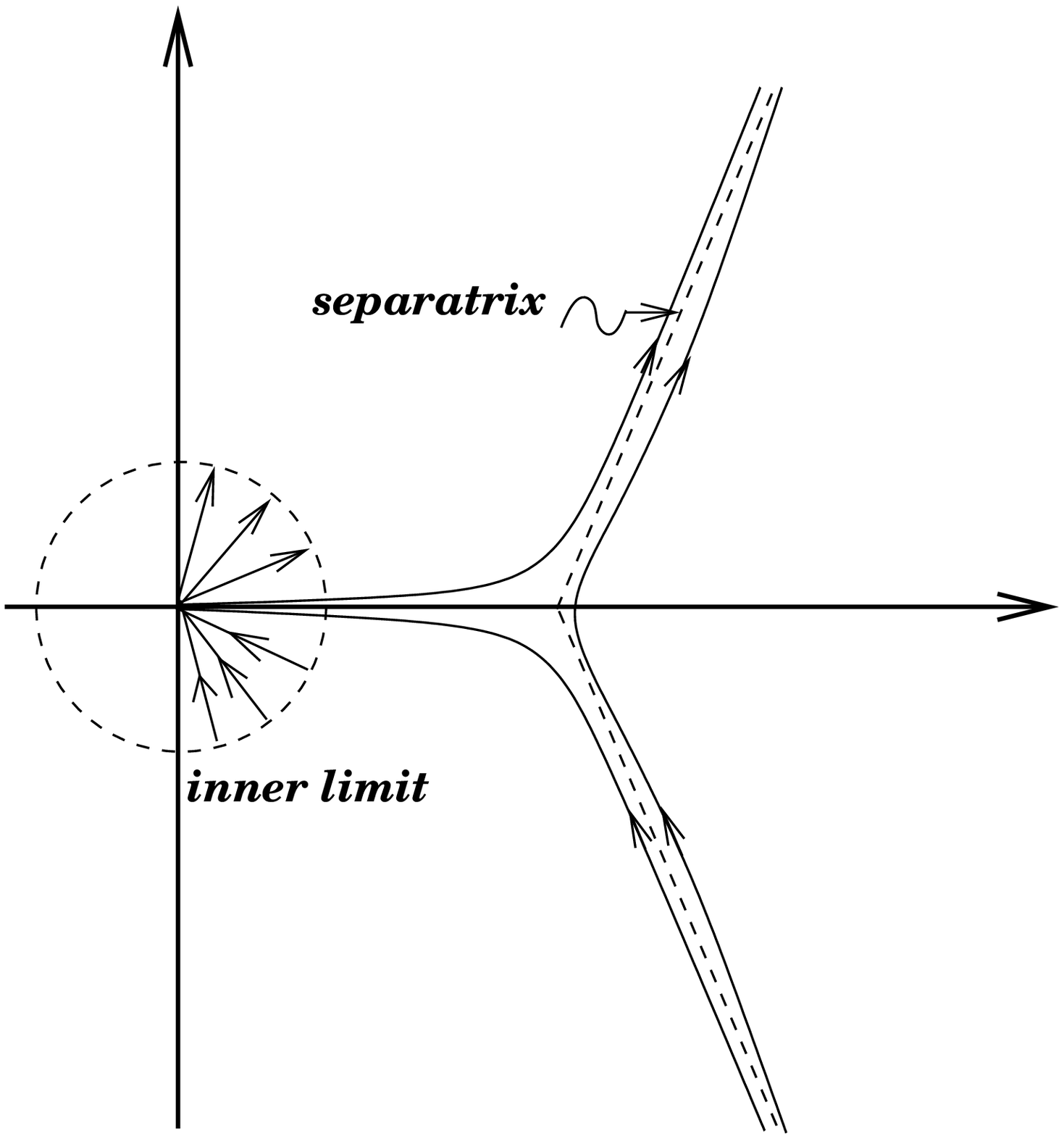}
\caption{Geometry of the magnetic field in a collapsing magnetized
cloud. The separatrix ({\it dashed curve}) separates field lines which
have been pulled into an accreting protostar (at the origin of the
coordinate system) from field lines tied to matter in the infalling
envelope. A {\it dashed circle} indicates the inner region with
nondimensional radius $x\ll 1$ studied in this paper, where magnetic
field lines are asymptotically radial.}
\label{fig1}
\end{figure}

The geometry of the collapse is shown in Figure~\ref{fig1}. A
separatrix indicated in dashed lines separates field lines which have
been pulled into the origin by the inflow from field lines tied to
matter in the infalling envelope (see, e.g., Figures 4, 6 and 7 of
A03a). We will study only the region inside the separatrix surface that
is close to the origin (within the dotted circle in
Figure~\ref{fig1}).

\section{Self-similar variables}

Since the singular isothermal toroids defining the pivotal $t$ = 0
states are self-similar, we look for collapse solutions depending
only on the similarity coordinate ${\bf x}$, defined by
\be
{\bf r}=at {\bf x}.
\label{defr}
\ee
We introduce appropriate nondimensional (reduced) variables
defined as
\be
\rho({\bf r},t)={1\over 4\pi Gt^2}\alpha({\bf x}),
\ee
\be
{\bf u}({\bf r},t)=a{\bf v}({\bf x}),
\ee
\be
V({\bf r},t)=a^2\Psi({\bf x}),
\ee
\be
{\bf B}({\bf r},t)={a\over G^{1/2}t}{\bf b}({\bf x}).
\label{defB}
\ee

Inserting the definitions (\ref{defr})--(\ref{defB}) into
eq.~(\ref{cont})--(\ref{ind}), we obtain the set of
nondimensional equations governing the self-similar collapse,
\be
({\bf v}-{\bf x})\cdot\nabla\alpha+\alpha\nabla\cdot{\bf
v}=2\alpha,
\label{sscont}
\ee
\be
-({\bf x} \cdot \nabla){\bf v}
+ (\nabla \times {\bf v}) \times {\bf v} +{1 \over 2} \nabla |{\bf
v}|^2 = -{1 \over \alpha} \nabla \alpha -\nabla \Psi + {1 \over
\alpha} (\nabla \times {\bf b}) \times {\bf b},
\ee
\be
-({\bf x}\cdot \nabla ) {\bf b} = {\bf b} + \nabla \times ({\bf v} \times
{\bf b}),
\label{ssind}
\ee
\be
\nabla \cdot {\bf b} = 0,
\label{nomono}
\ee
\be
\nabla^2\Psi=\alpha.
\ee

\section{The inner limit}

In the limit $|{\bf x}|\ll 1$ we approximate
for any vector or scalar variable ${\bf f}$,
\be
{\bf f} \sim ({\bf x} \cdot \nabla ) {\bf f}
\ll ({\bf v} \cdot \nabla ) {\bf f} .
\ee
With these approximations, eq.~(\ref{sscont})--(\ref{ssind}) become
\be
\nabla\cdot (\alpha {\bf v} )=0,
\label{slcont}
\ee
\be
(\nabla \times {\bf v}) \times {\bf v}
+{1 \over 2} \nabla |{\bf v}|^2 =  -{1 \over \alpha} \nabla \alpha
-\nabla \Psi + {1 \over \alpha} (\nabla \times {\bf b}) \times {\bf b},
\label{slforce}
\ee
\be
 \nabla \times ({\bf v} \times {\bf b}) = 0 .
\label{slind}
\ee
These equations for the self-similar variables in the inner region are
formally equivalent to the dimensional equations of steady, ideal MHD
flow.  The time-dependent self-similar flow becomes quasi-steady when
the time it takes the flow to cross the inner region becomes small
compared to the evolutionary time governing changes of the infall
envelope that feeds matter and field into the inner regions.  The
latter condition holds independent of the starting assumption of
self-similarity of the overall collapse (see e.g. Galli \& Shu~1993b),
and therefore our results have greater generality than its formal
derivation here. In the steady-flow approximation, we may profitably
introduce a streamline formalism (e.g, Shafranov 1966; Shu al.~1994).

In the following we adopt spherical coordinates $(x, \theta, \varphi)$, and
assume that the system is axisymmetric ($\partial / \partial \varphi = 0)$.
Then eq.~(\ref{slcont}) is satisfied introducing a stream function
$\psi({\bf x})$ such that
\be
\alpha{\bf v}_p=\nabla\times
\left[{\psi\over x\sin\theta}\hat{\bf e}_\varphi\right],
\ee
where the components of the poloidal velocity ${\bf v}_p$  are
\be
v_x={1\over \alpha x^2\sin\theta}{\partial\psi\over\partial\theta} \qquad
{\rm and} \quad
v_\theta=-{1\over \alpha x\sin\theta}{\partial\psi\over\partial x}.
\label{vrvt}
\ee
Reflection symmetry requires $\psi$ to be an odd function
of $\theta$ with respect to $\theta=\pi/2$, therefore
\be
\psi\left(x,{\pi\over 2}\right)=0.
\ee
Imposing that the integral over solid angle of the mass flow $-r^2\rho u_r$
gives the accretion rate $\mdot=m_0(1+H_0)a^3/G$, where $m_0=0.975$ is the
reduced mass at the origin (Li \& Shu 1997), we obtain the condition
\be
\psi(x,0)= m_0(1+H_0).
\label{BC}
\ee
Thus, the value of $\psi/\psi(x,0)$ is the fraction of the accretion
rate carried by the streamlines from $-\psi$ to $\psi$.

Eq. (\ref{slind}) implies that ${\bf v} \parallel {\bf b}$.  We define
the function $\beta$, related to the mass loading of field lines, by
\be
{\bf b } = \beta \alpha {\bf v}.
\label{par}
\ee
Using this definition, the condition of no monopoles, eq. (\ref{nomono}),
and the equation of continuity, eq. (\ref{slcont}), one obtains
\be
{\bf v} \cdot \nabla \beta = 0,
\label{beta}
\ee
i.e., $\beta = \beta(\psi)$, is constant along streamlines.  We note
that $\beta$ is antisymmetric with respect to the equatorial plane
(by convention, $\beta$ negative corresponds to the upper hemisphere).

If we dot the force equation (\ref{slforce}) with ${\bf v}$ we obtain
Bernoulli's equation \be {1\over 2}|{\bf v}|^2+\ln\alpha + \Psi=H(\psi),
\label{bern} \ee where $H(\psi)$ is Bernoulli's function which is constant
on streamlines. Substituting this equation and using the definition of
$\beta$, eq.~(\ref{par}), in the equation of motion, eq. (\ref{slforce}),
we obtain after collecting terms,
\be
H^\prime\nabla\psi+
[\nabla\times(1-\alpha\beta^2){\bf v}
+\nabla\beta\times\beta\alpha{\bf v}] \times {\bf v}=0,
\label{slforce2}
\ee
where a prime denotes differentiation with respect to $\psi$.  We now
decompose this equation along $\hat{\bf e}_\varphi$ and $\nabla\psi$.

After some algebra, one can write the poloidal component of
eq.~(\ref{slforce2}) as
\be
\left(H^\prime +{\omega\over \alpha x\sin\theta}
- {v_\varphi \over x \sin\theta} j^\prime
-\alpha\beta\beta^\prime |{\bf v}|^2\right) \nabla \psi = 0,
\ee
where
\be
\omega \hat {\bf e}_\varphi \equiv
\nabla\times(1-\alpha\beta^2){\bf v}_p
\ee
is the total vorticity.
For nontrivial solutions of this equation to exist, we require
\be
H^\prime +{\omega\over \alpha x\sin\theta}
- {v_\varphi \over x \sin\theta} j^\prime
-\beta\beta^\prime\alpha |{\bf v}|^2 = 0,
\label{GS}
\ee
which is a form of the Grad-Shafranov (G-S) equation that expresses the
condition of force balance across field lines.  We can write the total
vorticity as
\be
\omega = - {1 \over x \sin\theta} \left[A \,
{\cal S}(\psi) -  2 \beta \beta^\prime |\nabla \psi|^2 -
{1 \over \alpha^2} \nabla \alpha \cdot \nabla \psi \right],
\ee
where ${\cal S}$ is the Stokes operator
\be
{\cal S}(\psi) = {\partial^2 \psi \over \partial x^2} +
{1 \over x^2} {\partial^2 \psi \over \partial \theta^2}
-{\cot\theta \over x^2} {\partial \psi \over \partial \theta} ,
\ee
and $A$ is the Alfv\`en discriminant
\be
A =  {1 - \alpha \beta^2 \over \alpha}.
\ee
Then, the G-S equation, (\ref{GS}), can be written as
\be
A {\cal S}(\psi)  =
\alpha H^\prime x^2 \sin^2\theta +
{1 \over \alpha^2} \nabla \alpha \cdot \nabla\psi
- {1 \over A } \, j j^\prime
+ \beta \beta^\prime |\nabla \psi|^2 .
\label{acro}
\ee

The toroidal component of equation (\ref{slforce2}) is
\be
{\bf v} \cdot ( \nabla j  + \alpha\beta v_\varphi \nabla \beta ) = 0,
\label{tor}
\ee
where we have defined the total (gas plus magnetic
field) specific angular momentum
\be
j\equiv x\sin\theta(1-\alpha\beta^2)v_\varphi.
\label{jpsi}
\ee
Since
$\beta $ is constant along streamlines by eq. (\ref{beta}),
eq.~(\ref{tor}) reduces to
\be
{\bf v} \cdot  \nabla j  = 0,
\label{jadv}
\ee
stating that $j = j(\psi)$ is constant along
streamlines. Thus, the toroidal velocity is given by
\be
v_\varphi
= {j(\psi) \over x\sin\theta(1-\alpha\beta^2)}.
\label{vjpsi}
\ee
Notice the formal singular point that arises when $\alpha\beta^2 = 1$,
which is the condition of the poloidal velocity reaching the Alfven
speed.  In fact, the smooth crossing of this surface (with a
continously varying $v_\varphi$) is made in a time-dependent way; as
will be shown in Sect.~5, the poloidal velocity is distinctly
subalfv\`enic ($\alpha\beta^2 \gg 1$) in the steady part of the flow.
Finally, Bernoulli's eq. (\ref{bern}) can be rewritten as
\be
|\nabla \psi|^2+ \left({j \over A} \right)^2 + 2 \alpha^2
x^2 \sin^2\theta \left(\ln\alpha + \Psi-H \right) = 0.
\label{alo}
\ee

{Thus, in the inner limit, the ideal MHD equations of the problem
(\ref{slcont})--(\ref{slind}), can be reduced to the G-S and
Bernoulli's equations (eq.~\ref{acro} and \ref{alo}, respectively).
These need to be solved for the stream function $\psi$ and the density
$\alpha$ (or the Alfv\`en discriminant $A$), given the functions
$H(\psi)$, $j(\psi)$, and $\beta(\psi)$, which are quantities conserved
along streamlines. Then, the toroidal components of the magnetic field
and the velocity can be obtained from eq. (\ref{par}) and (\ref{vjpsi}),
respectively.}

\section{Case of no rotation ($j=0$)}

We first consider the case of no rotation, and we look for a separable
solution in spherical coordinates in which all variables have a power-law
dependence on $x$.  One can show that the density and the stream function
must have forms that represent magnetically-modulated free-fall solutions:
\be
\alpha(x,\theta) = \left[{ m_0 (1+H_0) \over 2
}\right]^{1/2} x^{-3/2}Q(\theta),
\label{densj0}
\ee
and
\be
\psi(x,\theta) = m_0 (1+H_0) f(\theta),
\label{psij0}
\ee
where $Q(\theta)$ is normalized so that
\be
\int_0^{\pi/2} Q(\theta)\sin\theta\, d\theta=1.
\ee
The angular part of the stream function, $f(\theta)$, is normalized so
that $f(0)=1$ and $f(\pi/2)= 0$, i.e., the streamlines between those
falling in equatorially and along the pole carry the entire mass infall
rate of the upper hemisphere.

To find the unknown functions $Q(\theta)$ and $f(\theta)$ we need to
balance the highest order terms in eqs. (\ref{acro}) and (\ref{alo}). The
dominant terms that balance in the G-S equation, eq. (\ref{acro}),
are then
\be
-\beta^2 {\cal S}(\psi) \approx \beta \beta^\prime |\nabla
\psi|^2,
\label{balance_GS}
\ee
where the Alfv\`en discriminant, $A\approx
-\beta^2$, corresponds to highly subalfv\`enic poloidal flow.

In Bernoulli's equation the dominant terms that balance are the
poloidal kinetic energy and the gravitational potential energy.
With the gravitational potential of a point mass at the origin,
$\Psi = -m_0(1+H_0)/x$, we get
\be
|\nabla \psi|^2 \approx 2
m_0 (1+H_0) \alpha^2 x \sin^2\theta.
\label{balance_bern}
\ee
Substituting the definitions (\ref{densj0}) and (\ref{psij0}) in
this latter equation, one gets
\be
{d  f \over d \theta}
= - Q(\theta) \sin\theta.
\label{bernj0}
\ee
The G-S equation
(\ref{balance_GS})now  becomes a second-order, nonlinear, ordinary
differential equation for $f(\theta)$,
\be
{d^2 f \over d \theta^2} - \cot\theta {d f \over d \theta}
+{\beta^\prime \over \beta} \left({d f \over d
\theta}\right)^2 =0.
\label{balj0}
\ee
Combining the Bernoulli
equation (\ref{bernj0}) and the G-S equation (\ref{balj0}), one
obtains
\be
{d Q \over d f} = - {\beta^\prime \over \beta} Q,
\ee
which has the solution
\be
Q(f) = {\bar\beta \over \beta(f)},
\label{asol}
\ee
where $\bar\beta$ is an integration constant. Furthermore,  $f(\theta)$
has the parametric solution
\be \cos\theta(f) = {1 \over
\bar\beta} \int_0^{f} \beta(f)\, df,
\label{cossol}
\ee
where we have used the B.C. $f(\pi/2)=0$. Moreover, the B.C. $f(0)=1$,
implies that
\be
\bar \beta = \int_0^1 \beta(f)\, df.
\label{bbar}
\ee

The physical meaning of the above results is simple.
Equation~(\ref{cossol}) implies that the poloidal field lines spread so
as to distribute themselves uniformly in polar angle, i.e., the
magnetic configuration takes the shape of a {\it split monopole}, {
as anticipated by Galli \& Shu (1993a; see their Figure 2) and
confirmed later by Li \& Shu~(1997) for the collapse of an
infinitesimally thin disk.}
This happens because the field lines trapped in the
central mass point become so strong that the mass inflow, being highly
subalfv\`enic, is unable to bend them from the zeroth-order vacuum
field configuration.  In turn, the matter is constrained by field
freezing to flow along the radial field lines, and the density
distribution is that appropriate for free-fall along each radial path
(eq.~\ref{psij0}), but it departs from a homogenous angular
distribution in a manner that depends only on the initial mass-to-flux
loading (eq.~\ref{asol}).

Given the above interpretation, it is useful to define
the magnetic flux function
\be
\Phi ({\bf r},t)= {2 \pi a^3 t \over G^{1/2}} \phi({\bf x}),
\ee
from which we may obtain the poloidal field ${\bf B}$:
\be
{\bf B}=\nabla\times\left({\Phi\over 2\pi r\sin\theta}\hat{\bf e}_\varphi\right).
\ee
The components of the poloidal magnetic field, in nondimensional units,
are given by
\be
b_x={1\over x^2\sin\theta}{\partial\phi\over\partial\theta},\qquad
{\rm and} \quad
b_\theta=-{1\over x\sin\theta}{\partial\phi\over\partial x}.
\label{defb}
\ee
Comparing these definitions with equations~(\ref{vrvt}), and using the
definition of $\beta$, eq.~(\ref{par}), one sees that there is a simple
relation between the flux and the stream function given by
\be
d\phi = \beta\, d\psi
\label{dphidpsi}
\ee
Thus, the
parametric equation for $\psi$, eq.~(\ref{cossol}), simply states
that \be
\phi = \phi_\star (1 -\cos\theta),
\ee
where
$\phi_\star = \phi(\pi/2)$ (just above the midplane) is the
nondimensional magnetic flux trapped in the central source. In fact,
$\phi_\star$ labels the flux of the separatrix (see
Figure~\ref{fig1}). Moreover, the normalization eq.~(\ref{bbar}) and
eq.~(\ref{dphidpsi}) give
\be
\bar\beta= -{\phi_\star\over m_0(1+H_0)}.
\label{Csol}
\ee
Since every field line that lies interior to the separatrix threads the
mass point at the center, one has the immediate identification that
\be
\lambda_\star = -\bar\beta^{-1}={m_0 (1+H_0)\over \phi_\star},
\label{lambdastar}
\ee
where $\lambda_\star$ is the nondimensional mass-to-flux ratio of the central star defined as
\be
\lambda_\star
\equiv { 2 \pi G^{1/2} M_\star \over \Phi(r, \pi/2)}.
\ee

To summarize, the solution for small $x$ consists of streamlines
which are radial but non-uniformly distributed in $\theta$.  The
$\theta$ dependence of both $\psi$ and $\alpha$ is determined by
the mass loading of field lines represented by $\beta(\psi)$,
assumed to be known from the matching to the outer collapse
solution.  According to eqs.  (\ref{vrvt}) and (\ref{bernj0}), the
radial velocity is just free-fall, i.e.,
\be
v_x = - \left[{ 2 m_0 (1 + H_0) \over x} \right]^{1/2}.
\label{vx}
\ee
By eq. (\ref{par}), the magnetic field is that of a split
monopole at the the origin,
\be
b_x = {\phi_\star \over x^2},
\label{bx}
\ee
as inferred by A03a,b.
In physical units, the magnetic field of the split monopole is
\be
B_r=\phi_\star{a^3 t\over G^{1/2}r^2},
\ee
and is maintained by a current sheet in the equatorial plane, with current
density in the azimuthal direction
\be
J_\varphi={c B_r\over 2\pi r}\delta(\theta-\pi/2).
\ee

Since the radial velocity increases as $v_x \propto x^{-1/2}$,
while the (nondimensional) Alfv\`en speed increases much faster,
$v_A= b_x / \alpha^{1/2} \propto x^{-5/4}$, the flow asymptotically
becomes highly subalfv\`enic, consistent with our assumption for $A$
in eq. (\ref{balance_GS}).

\section{Case with rotation ($j\neq 0$).}

In the case $j=0$, the balance across and along streamlines is achieved by
terms of order $x^{-2}$ (see eq.~\ref{balance_GS} and \ref{balance_bern}).
Multiplying eq.~(\ref{acro}) by $A$, and eq.~(\ref{alo}) by $A^2$,
one can show that the highest-order balance of the G-S and Bernoulli
equations for $j\neq 0$ is unchanged. Thus, the non-rotating solution for
the density $\alpha$ and the radial velocity and magnetic field, $v_x$
and $b_x$, derived in the previous Section remains valid for $x\ll 1$,
for any value of $j$. This is consistent with the findings of A03b that
the isodensity contours and poloidal field (stream) lines are the same to
zeroth approximation in the rotating and non-rotating case for small $x$.

Using eq. (\ref{jadv}) together with the equation of continuity,
eq. (\ref{slcont}), the conservation of total angular momentum,
$\nabla \cdot (\alpha {\bf v} j )=0$, can be written as \be \nabla
\cdot (\alpha {\bf v}_p j_g) = \nabla \cdot ({\bf b}_p b_\varphi x
\sin\theta) \ee where the gas specific angular momentum is $j_g =
x \sin\theta v_\varphi$. This equation can be integrated over the
upper hemisphere as
\be
\int \alpha j_g {\bf v}_p \cdot d{\bf S}=
\int x \sin\theta b_\varphi {\bf b}_p \cdot d{\bf S},
\ee
where $d {\bf S} = 2 \pi x^2 \sin\theta d\theta \, \hat {\bf e}_x$. The
left hand side is rate of change of the gas specific angular momentum
and the right hand side is the magnetic torque. Given our inner
solution, the integral on the left hand side is $\propto x v_\varphi$
and the magnetic torque is $\propto x^{-1/2} v_\varphi$.  Thus, the
magnetic torque eventually dominates, decreasing the net gas angular
momentum and forcing the azimuthal velocity to deviate from the
$v_\varphi\propto x^{-1}$, behavior valid in the absence of magnetic
torques.  As discussed above, for small $x$  the flow becomes
subalfv\`enic ($\alpha \beta^2 \gg 1$). Thus, eq. (\ref{vjpsi}) for the
azimuthal velocity, together with eq. (\ref{densj0}), implies that
along each streamline $v_\varphi \propto x^{1/2}$, i.e., the azimuthal
velocity decreases to zero at the origin. The long lever arm associated
with the divergently strong split monopole field yields highly
efficient magnetic-braking, causing the gas to spiral into the central
star without forming a centrifugally supported disk, as suggested by
A03b. 

{Finally, from eq.~(\ref{par}), one can obtain the toroidal
magnetic field component $b_\varphi \propto x^{-1}$, that increases
with decreasing radius more slowly than the poloidal component $b_x$
(eq. (\ref{bx})).  Thus, the winding of the field
($b_\varphi/b_x\propto x$) goes to zero as $x\rightarrow 0$, where the
field geometry is dominated by the split monopole at the origin.  Thus,
in our treatment, reconnection of the poloidal magnetic field because
of the sheet current that prevails in the magnetic midplane is more
important than reconnection of the toroidal component that results
because of the spinup of inflowing matter. In realistic circumstances,
this state of affairs may change after the poloidal field is nearly
entirely dissipated and a centrifugal disk is formed that can wind up
relatively weak fields into a predominantly toroidal configuration.}

\begin{figure}
\plotone{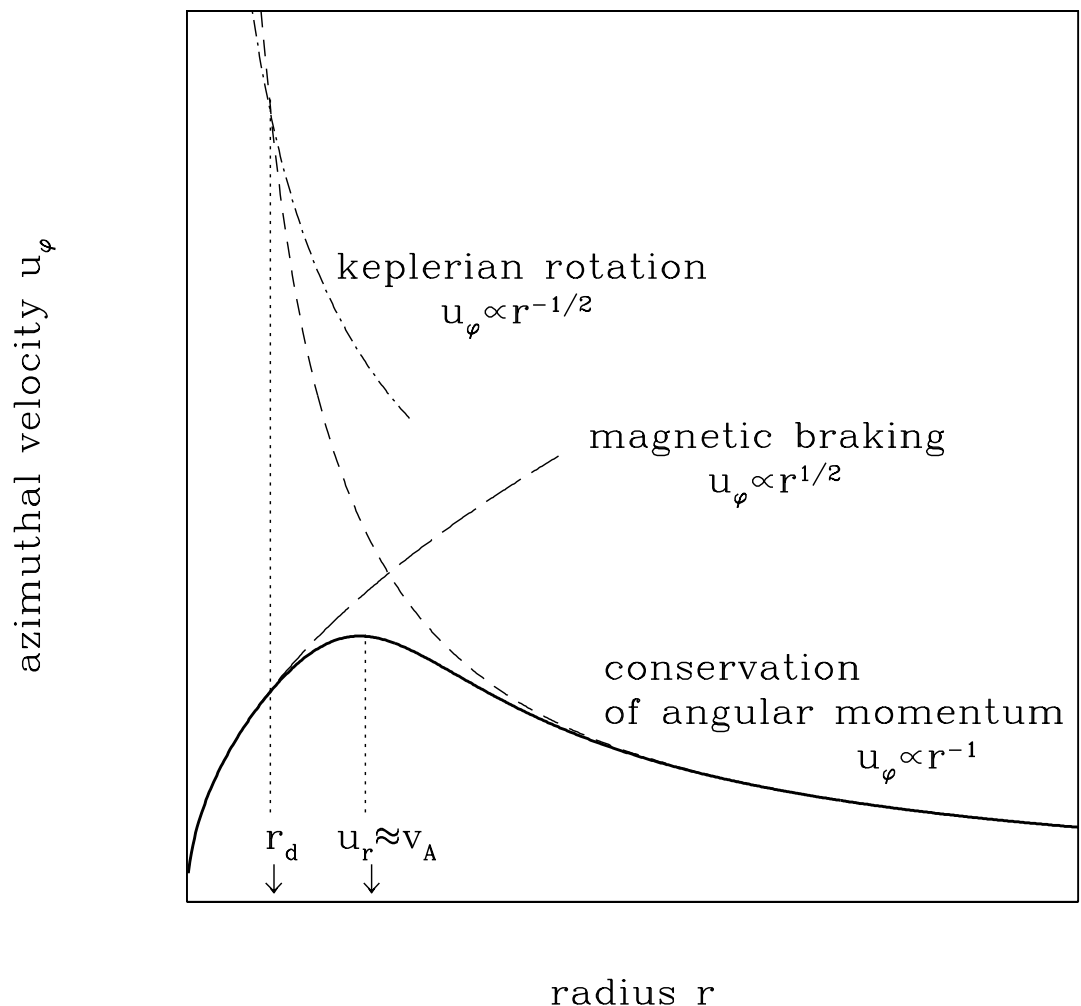}
\caption{Schematic behavior of the azimuthal velocity $u_\varphi$ ({\it
solid curve}) in the equatorial plane of a collapsing cloud as function
of the distance $r$ from the central protostar. In the absence of
magnetic torques, a centrifugally supported disk is formed inside a
radius $r_d$, where the azimuthal velocity increasing as $r^{-1}$
becomes equal to the keplerian velocity around the protostar ($\propto
r^{-1/2}$). In contrast, the magnetic torque associated with the split
monopole field of the central protostar reduces the angular momentum of
the infalling gas, constraining the azimuthal velocity to decrease as
$r^{1/2}$ at small radii. Magnetic braking becomes dominant over
angular momentum conservation when the infall velocity $u_r$ becomes
smaller than the local Alfv\`en speed $v_{\rm A}$.}
\label{fig2}
\end{figure}

Figure~\ref{fig2} summarizes the results of this Section and
illustrates qualitatively the behavior of the azimuthal velocity in the
equatorial plane of a collapsing magnetized cloud.  {We stress that
different choices of the mass-loading of field lines specified by the
function $\beta(\psi)$ would merely change the details of when and
where the maximum rotational velocity is reached in the inflow
solution, not the fact that the inflowing matter eventually loses all
its specific angular momentum.  In the next section, adopting a
specific collapse model, we will show quantitatively that the region
where magnetic braking by the split monopole becomes dominant over
angular momentum conservation is generally larger than the size of a
centrifugally supported disk formed in the absence of magnetic
torques.}

\section{Matching to the outer collapse solution}

We can match our inner collapse solution $x \ll 1$ to the outer collapse
solution via the mass-loading function, $\beta(\psi)$, and the specific
total angular momentum, $j(\psi)$. We use the simulations of A03a to
perform this asymptotic matching. The exercise is nontrivial since
collapse simulations with very high grid resolution are required to
provide the ``inner limit of the outer solution,'' onto which we wish
to match the ``outer limit of the inner solution.'' For our purposes,
there is a further problem with the boundary condition imposed by Allen
and co-workers on the radial field, namely $b_x = 0 $ at $\theta=\pi/2$.
This boundary condition, appropriate to the outside of the separatrix,
is in conflict with the magnetic field configuration inside of the
separatrix (that becomes the split monopole solution at small $x$)
depicted in Figure~\ref{fig1}. In fact, unlike the analytic work,
it is numerically impossible to allow a current sheet at the equator
(where $b_x$ is discontinuous at the midplane) since this configuration
is unstable to magnetic reconnection via numerical diffusion (see, e.g.,
Figure 4 of Galli \& Shu 1993b). In spite of these drawbacks, we use the
numerical $\beta$ obtained from the simulations of A03a to apply to
our models, trusting the conservation principle associated with field
freezing to yield the correct mass-to-flux loading for all colatitudes
except close to the equator.

\begin{figure}
\plotone{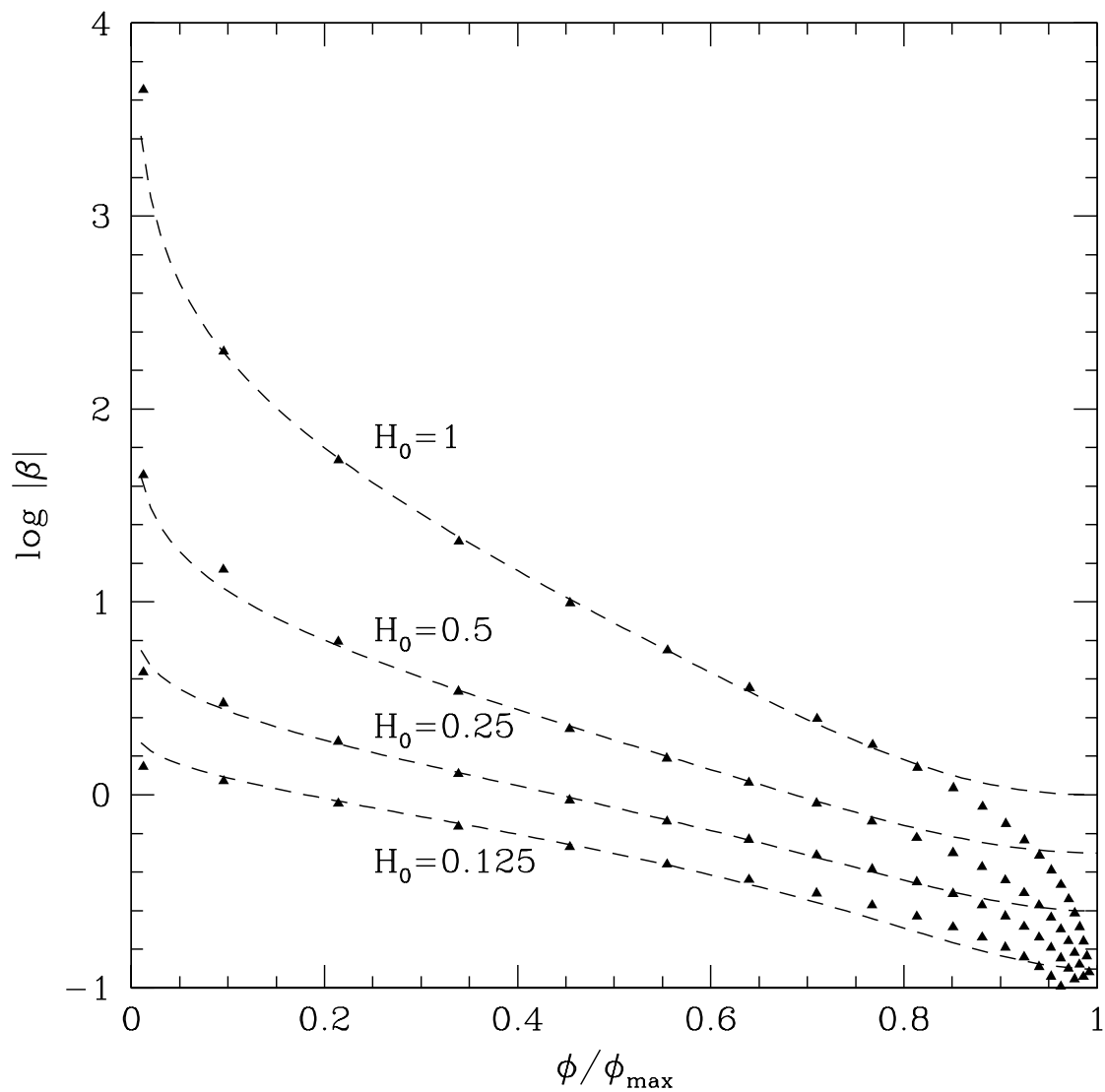}
\caption{
The function $\beta$ related to the mass loading of field lines
(eq.~\ref{par}), plotted as function of the normalized flux
$\phi/\phi_\star=1-\cos\theta$.  The {\it triangles} show the function
$\beta$ computed numerically by A03a for $H_0=0.125$, 0.25, 0.5 and 1;
the {\it dashed curves} are analytical fits given by eq.~(\ref{fits}).}
\label{fig3}
\end{figure}

In Figure~\ref{fig3} the logarithm of $|\beta|$ is plotted as a function of
the normalized flux, $\xi =\phi/\phi_\star$, for different
values of $H_0$ increasing from bottom to top. The points are the
results of the simulations. The broken lines are fits given by
\be
-\beta_{H_0} = b \exp(-c \xi) \xi^{-H_0}\left(1-\xi^4\right)^2 +
H_0,
\label{fits}
\ee
where $ b=H_0^{0.2} \exp(3.3 H_0)$ and $c=1.6+2.3 H_0$.  These broken
lines yield a good representation of the numerical data for all angles
except those close to the equator, $\xi =1$.  The numerical $\beta$ goes
to zero due to the boundary condition discussed above (see eq.~\ref{par}).
The analytic solution requires $\beta$ to be finite and $\beta^\prime=0 $
at the midplane; otherwise the density gradient will become discontinuous.

Note that the function $\beta$ diverges at the pole ($\theta=0$)
because in the starting toroids the density there goes to zero.
Finally, since the numerical simulations give $\beta(\phi)$, not
$\beta(\psi)$, we use the fact that $d \psi = \phi_\star \, d
\xi/\beta(\xi)$, to obtain from eq.~\ref{dphidpsi}) and
(\ref{lambdastar}) \be \lambda_\star=-\int_0^1 {d\xi\over\beta(\xi)}.
\ee The numerical values of $\lambda_\star$ can be determined only
approximately, owing to the effects of the boundary condition at the
midplane on the numerical values of $\beta$. Roughly, we obtain
$\lambda_\star\approx 3$--4 for $H_0=0.125$, $\lambda_\star\approx
2.4$--2.7 for $H_0=0.25$, $\lambda_\star\approx 1.2$--1.7 for
$H_0=0.5$, and $\lambda_\star=1.0$--1.3 for $H_0=1$.  These values are
closer to the mass-to-flux ratio $\lambda_r$ than they are to $\lambda$
as tabulated in Table 1 of Li \& Shu (1996).  The former applies for
the initial state of an isopedic toroid when one integrates for the
mass along flux tubes for given $H_0$ only to a spherical surface of
radius $r$, whereas the latter applies if one integrates all the way
out to $\infty$. For given trapped flux, we can expect mass
accumulation to behave in the latter case only in the case of complete
flattening, $H_0 \rightarrow \infty$.  For more modest values of $H_0$,
we can expect a more nearly spherical reach for both mass and flux when
the configuration collapses, yielding the closer correspondence of
$\lambda_\star$ to $\lambda_r$.

\begin{figure}
\plotone{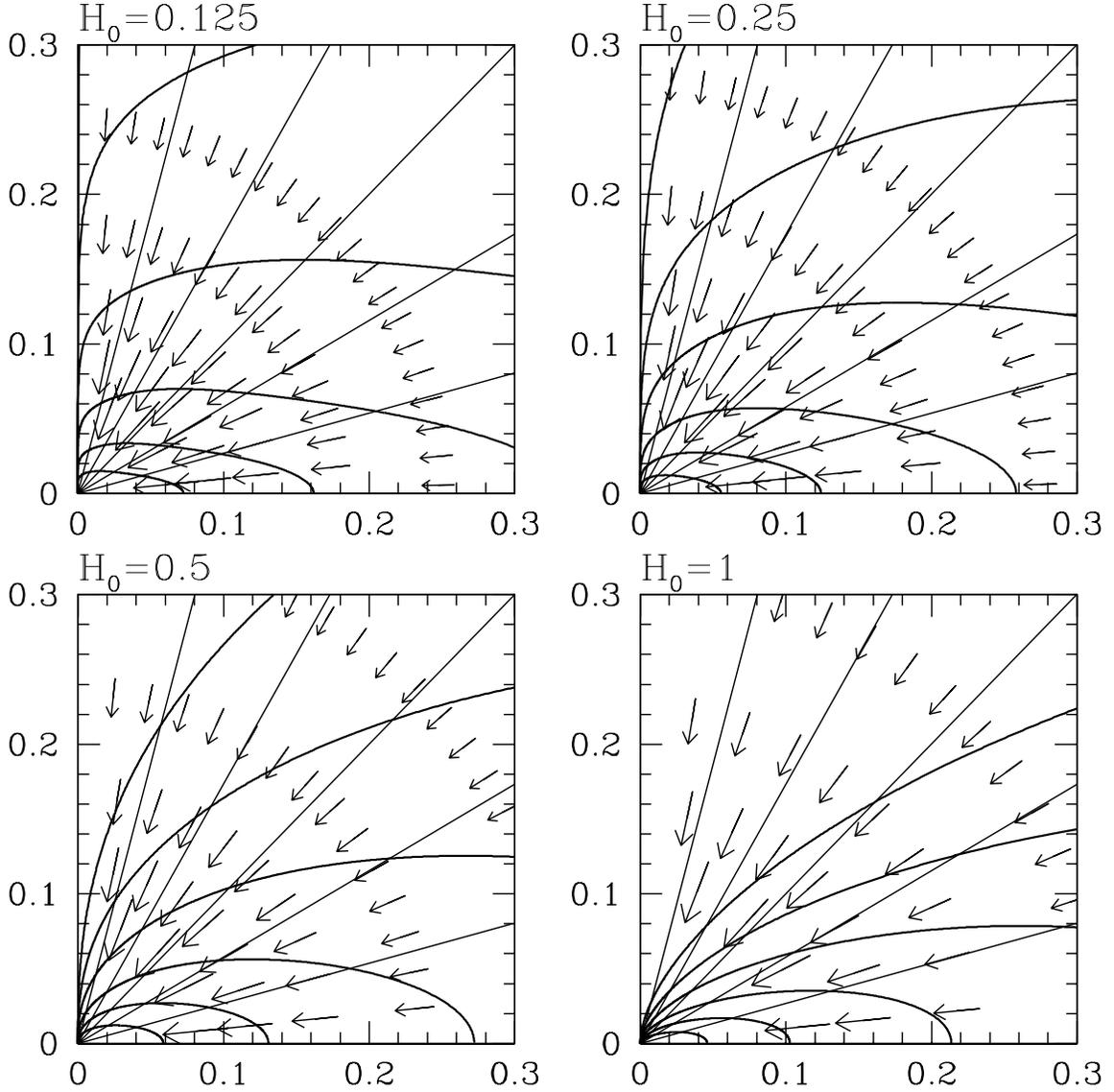}
\caption{Inner collapse solutions (valid asymptotically for $x\ll 1$)
obtained for the values of the parameter $H_0$ and the functions $\beta$
shown in Figure~\ref{fig3}. The horizontal and vertical axis in each panel
are the cylindrical self-similar coordinates, $\varpi = x \sin\theta$
and $z=x\cos\theta$. The {\it heavy solid contours} in each panel are
isodensity contours for the levels $\alpha= 100, 30, 10, 3, 1$ and 0.3. The {\it
thin solid lines} are the magnetic field lines (which coincide with the
streamlines). The {\it arrows} show the velocity field at different radii,
logarithmically spaced in the velocity interval $[2.5, 6]$.}
\label{fig4}
\end{figure}

In contrast, T~Tauri stars have observed values of
$\lambda_\star\approx 10^3$--$10^4$ (see e.g. Johns-Krull et al.~2004),
implying that the assumption of field-freezing must break down at some
point in the star formation process (Shu et al.~2006). As expected,
according to our ideal MHD solution the magnetic field trapped in the
central protostar exceeds observed stellar fields by three-four orders
of magnitude.  The four panels of Figure~\ref{fig4} show our inner
collapse solutions in self-similar coordinates for the four values of
$H_0$ given above.  The heavy solid contours in each panel are
isodensity contours and the radial lines are the magnetic field lines
(which coincide with the streamlines). The arrows show the velocity
field at different radii.  The dynamic pseudodisks that result may bear
comparison with observations of young protostars before they have
developed full disks (Hogerheijde 2004).

{Quantitatively, the condition for subalfv\`enic radial flow
$\alpha\beta^2 >1$, necessary for strong magnetic braking according to
the results described in the previous section, is valid in a region $x
< x_A(\theta)$ where $x_A$ is the nondimensional Alfv\`en radius.  With
our asymptotic expressions for $v_x$ and $b_x$ (eq.~\ref{vx} and
\ref{bx}) and our fitting formula for $\beta$ (eq.~\ref{fits}), it is
possible to estimate the value of $x_A$ on the midplane
($\theta=\pi/2$), obtaining $x_A\approx 0.085$, 0.18, and 0.45 for
$H_0=0.125$, 0.25 and 0.5.  These values can be compared with the
expected disk radius for the collapse of a non-magnetized
Toomre-Hayashi toroid, $x_d\approx 0.25 v_0^2$, where $v_0$ is the
(uniform) rotation velocity in units of the isothermal sound speed $a$
(A03b). The magnitude of $v_0$ for molecular cloud cores ranges from
0.03 to 0.4, with typical value $v_0\approx 0.1$ (Goodman et al.~1993),
in which case $x_d\approx 0.0025$. Thus, $x_d \ll x_A$ even for weakly
magnetized clouds ($H_0=0.125$--0.5), and the formation of a
centrifugally supported disk cannot take place.  Even for the
weakly-magnetized, fast-rotating cloud model with $v_0=0.5$ and
$H_0=0.25$ considered by A03b (see their Fig.~7), the radius of the
disk that would be formed in the absence of magnetic torques,
$x_d\approx 0.063$, is about 3 times smaller than the Alfv\`en radius,
$x_A\approx 0.18$, explaining why A03b did not observe the formation a
centrifugal disk even in this extreme model.}

\section{Comparison with previous work}

{The process of magnetic braking has been usually considered in the
context of the quasi-static evolution of a molecular cloud core (e.g.
Gillis, Mestel \& Paris~1974,1979; Mouschovias \& Paleologou~1979).
Recent semi-analytical calculations and numerical simulations have
extended these works to follow the increase of density towards the
formation of an optically thick core (Tomisaka~2002, Machida et
al.~2005a,b) and through the formation of a point-mass into the
protostellar accretion phase (Li \& Shu~1997; Ciolek \& K\"onigl~1998;
Contopoulos, Ciolek \& K\"onigl~1998; A03a,b). Krasnopolsky \& K\"onigl
(2002) extended the semi-analytical calculations by Contopoulos et al.
(1998) to study the collapse of magnetized, rotating, thin disks
surrounded by an external medium with specified physical properties.
They found that the decoupling of matter and magnetic field (by
ambipolar diffusion) is carried by an outward propagating shock, as
first suggested by Li \& McKee (1996), and the infalling rotating
material is halted by a centrifugal shock inside which a centrifugal
disk is established. The formation and radial extent of the centrifugal
disk depend on two main parameters: the ratio $\delta$ of the azimuthal
and vertical components of the magnetic field at the disk surface and
the Alfven speed in the external medium, both controlling the
importance of magnetic braking during the collapse. In the ideal MHD
case (no ambipolar diffusion), for models with a large winding of the
magnetic field ($\delta=10$) they obtain complete braking, i.e. no
centrifugal disk is formed. As in our case, the horizontal flow
approaches free-fall onto the central mass and the magnetic field of
the protostar is that of a split monopole at the origin.  However, our
model differs from theirs in that our collapsing cores are 3-D with
axial symmetry and we do not include an external medium. The magnetic
braking is due to the field lines that connect the inner region, $x \ll
1$, with the rest of the collapsing cloud and thus the magnetic field
winding and magnetic braking are calculated self-consistently (through
the functions $j$ and $\beta$).  In addition, we find that the radial
component of the field largely dominates over the azimuthal component
at small radii, so the winding of the field is never as severe as
assumed by Krasnopolsky \& K\"onigl (2002).  }

\section{Summary and conclusions}

We have examined the properties of the inner regions of an ideally
magnetized, isothermal, rotating cloud undergoing gravitational
collapse. The density distribution and the poloidal velocity field in the collapse
solution approach free-fall forms, while the trapped magnetic field
rooted in the central protostar acquires a split monopole
configuration.  An analytic solution can be obtained for the central
regions if numerical simulations are used to obtain the loading
function that determines how much mass is coming down a given field
line. {For illustrative purposes, we have adopted the mass loading
function $\beta$ from the numerical simulations of A03a. We note,
however, that the properties of our collapse solution discussed in
Sect.~5 and Sect.~6 are general, and do not depend on these particular
numerical simulations.}

The long lever arm associated with the strong field is able to brake
the rotation of the infalling gas so efficiently that the azimuthal
velocity goes to zero at the origin.  In other words, the infalling gas
spirals directly into the center without the formation of a
centrifugally supported disk.  {Although this conclusion rigorously
holds only for the particular class of collapse models considered in
this paper, the assumption of self-similarity helps to brings out the
dependence of our results on the fundamental physical parameters of the
problem, like the cloud's mass-to-flux ratio and rotation rate,
suggesting a broader validity of our main conclusions.  Thus, the
catastrophic magnetic braking discussed in Sect.~6, for example, does
not depend on the particular choice of the outer collapse solution but
follows directly from the ideal MHD equations.} This work therefore
proves that magnetic field dissipation is a crucial ingredient to allow
the formation of the circumstellar disks observed around young stars.
It also yields departure points for calculations which include outflow
in the presence of inflow (Lizano et al.~2006, in preparation) or
modifications introduced by the effects of finite resistivity (Shu et
al.~2006).

\acknowledgements

DG and SL acknowledge financial support from the Theoretical
Institute for Advanced Research in Astrophysics (TIARA),
CRyA/UNAM, DGAPA/UNAM and CONACyT (Mexico), and INAF-Osservatorio
Astrofisico di Arcetri (Italy), where part of the research
presented in this paper was done. The authors are also grateful to
members and staff of these institutions for warm hospitality.  The
research work of FS and AA in Taiwan is supported by the grant
NSC92-2112-M-001-062.

\clearpage

\end{document}